\documentstyle{aipproc}
\begin{document}

\title{Synchronization of Chaotic Systems by Common Random Forcing}

\author{Ra\'{u}l Toral$^{1,2}$, Claudio R. Mirasso$^{2}$,
E. Hern\'{a}ndez-Garc\'{\i}a$^{1,2}$ and Oreste Piro$^{1,2}$}

\address{(1)Instituto Mediterr\'aneo de Estudios Avanzados,
IMEDEA\footnote{URL: http://www.imedea.uib.es/PhysDept}, CSIC-UIB.\\
(2) Departament de F\'{\i}sica, Universitat de les Illes Balears.\\
07071-Palma de Mallorca, Spain}
\date{January 31st 1999}
\maketitle

\noindent to appear in {\sl Unsolved Problems of Noise
and Fluctuations}: UPoN'99, Proceedings of the 2nd International
Conference in Adelaide (Australia). Edited by D. Abbot. American
Institute of Physics, (Melville NY, 2000).\\ Version with
postscript figures included available from \\
{\tt
http://www.imedea.uib.es/PhysDept/publicationsDB/date.html}
\begin{abstract}
We show two examples of noise--induced synchronization. We study a
1-d map and the Lorenz systems, both in the chaotic region.
For each system we give numerical evidence that
the addition of a (common) random noise, of large enough intensity,
to different trajectories which start from different initial
conditions, leads eventually to the perfect synchronization of the
trajectories. The largest Lyapunov exponent becomes negative due to
the presence of the noise terms.
\end{abstract}

\section{Introduction}

The issue of whether chaotic systems can be synchronized by common
random noise sources has attracted much attention
recently\cite{mar94,mal96,san97,min98,per99,ali97}. It has been reported
that for some chaotic maps, the introduction of the same (additive)
noise in independent copies of the same map could lead (for large enough
noise intensity) to a collapse onto the same trajectory,
independently of the
initial condition assigned to each of the copies\cite{mar94}.
This synchronization of chaotic systems by the addition of random terms
is a remarkable and counterintuitive effect of noise.
Nowadays, some contradictory results exist for the existence of this
phenomenon of noise--induced synchronization. It is purpose of this
paper to give explicit examples in which it is shown that one can
indeed obtain such a synchronization. Moreover, the examples open
the possibility to obtain such a synchronization in electronic circuits,
hence suggesting that noise-induced synchronization of chaotic circuits
can indeed be used for encryption purposes.

Although the issue of which is the effect of noise in chaotic systems
was considered at the beginning of the 80's\cite{mt83}, to our knowledge
the first atempt to synchronize two chaotic systems by using the
same noise signal was considered by
Maritan and Banavar\cite{mar94}. These authors
analysed the logistic map in the presence of noise:
\begin{equation}
x_{n+1}=4x_n(1-x_n)+\xi_{n}
\label{map}
\end{equation}
where $\xi _{n}$ is the noise term, considered to be uniformly
distributed in a symmetric interval $[-W,+W]$. They showed that, if
$W$ was large enough (i.e. for a large noise intensity) two different
trajectories which started with different initial conditions but used
otherwise the same sequence of random numbers, would eventually
coincide into the same trajectory. This result was heavily critisized
by Pikovsky\cite{pik94} who argued that two systems can synchronize
only if the largest Lyapunov exponent is negative. He then shows that
the largest
Lyapunov exponent of the logistic map in the presence of noise is
always positive and concludes that the
synchronization is, in fact, a numerical effect
of lack of precision of the calculation. Furthermore, Malescio\cite{mal96}
pointed out that the noise used to simulate Eq.(\ref{map}) in \cite{mar94}
was not really
symmetric. This is because the requirement $x_n \in (0,1),~\forall n$, actually
leads to discard those values for the random number $\xi_n$, which do not
fulfill such condition. The average value of the random numbers which
have been accepted is different from zero, hence
producing an effective {\sl biased} noise, i.e. one which does not have
zero mean. The introduction of a non-zero mean noise means that we are
altering essentially the properties of the deterministic map.

Noise induced synchronization has been since studied for other chaotic systems
such as the Lorenz model\cite{mar94,mal96} and the Chua
circuit\cite{san97,per99}.
Synchronization of trajectories starting with different initial conditions
but using otherwise the same sequence of random numbers
was observed in the numerical integration of a Lorenz system in the
presence of a noise distributed uniformly in the interval $[0,W_L]$, i.e.
again a noise which does not have a mean of zero\cite{mar94}. Other detailed
studies\cite{mal96} also conclude that it
is not possible to synchronize trajectories in a Lorenz system by adding an
unbiased noise.
Similarly, the studies of the Chua circuit always conclude that
a biased noise is needed for synchronization\cite{per99}.
Therefore a widespread belief exists that it is not possible to
synchronize two chaotic systems by injecting the same noisy signal to
both of them. However,
in this paper we give numerical evidence that
it is possible to synchronize
two chaotic systems by the addition of a common noise which is Gaussian
distributed and not biased.
We analyse
specifically a 1-d map and the Lorenz system, both in the chaotic region.
The necessary
criterion introduced in ref.\cite{pik94} is fully confirmed and some
heuristic arguments are given about the general validity of our results.
Finally, we conclude with some open questions relating the general
validity of our results.
\section{Results}
The first example is that of the map:
\begin{equation}
x_{n+1}= f(x_n)+\epsilon \xi_n
\label{eq:2}
\end{equation}
where $\xi_n$ is a set of uncorrelated Gaussian variables
of zero mean and variance 1. We use explicitely
\begin{equation}
f(x)=\exp\left[-\left(\frac{x-0.5}{\omega}\right)^2\right]
\label{eq:3}
\end{equation}
We plot in Fig.(1a) the bifurcation diagram of this map.  In the
noiseless case, we can see the typical windows in which
the system behaves chaotically. The associated Lyapunov exponent,
$\lambda$, in these regions is positive. For instance, for $\omega=0.3$
(the case we will be considereing throughout the paper) it is
$\lambda \approx 0.53$. In Fig.(1b) we observe that the Lyapunov
exponent becomes negative for most values of $\omega$ for large enough
noise level $\epsilon$. Again for $\omega=0.3$ and now for $\epsilon=0.2$ it
is $\lambda=-0.17$.

For the noiseless case, it is $\lambda >0$ and trajectories starting
with differential initial conditions, obviously, remain different for
all the iteration steps, see Fig.(2a). However,
when moderated levels of noise ($\epsilon \ge 0.1$) are used, $\lambda$
becomes negative and
trajectories starting with different initial conditions, but using the
same sequence of random numbers, synchronize perfectly, see Fig.(2b).

According to \cite{pik94},
convergence of trajectories to the same one, or loss of memory of the
initial condition, can be stated as {\sl negativity of the Lyapunov
exponent}. The Lyapunov exponent of a map $ x_{n+1}=F(x_n)$
is defined as
\begin{equation}
\lambda =\lim_{N\rightarrow\infty }\frac{1}{N}\sum_{i=1}^N \ln|F'(x_i)|
\end{equation}
It is the average of (the logarithm of the absolute value of) the
successive slopes $F'$ found by the trajectory.
Slopes in $[-1,1]$ contribute to $\lambda$ with
negative values, indicating trajectory convergence. Larger or
smaller slopes contribute with positive values, indicating trajectory
divergence.
Since the deterministic and noisy maps satisfy $F'=f'$  one
is tempted to conclude that the Lyapunov exponent is not modified by the
presence of noise. However, there is noise-dependence through the
trajectory values
${x_i}$, $i=1,...,N$. In the absence of noise, $\lambda$ is positive,
indicating trajectory separation.
When synchronization is observed, the Lyapunov exponent becomes negative,
as required by the argument in \cite{pik94}.

By using the
definition of the {\sl invariant measure on the attractor}, or {\sl
stationary probability distribution} $P_{st}(x)$, the Lyapunov exponent
can be calculated also as
\begin{equation}
\lambda=\left< \log|F'(x)| \right> = \left< \log|f'(x)| \right> \equiv
\int P_{st}(x) \log|f'(x)| dx
\label{lyapunov}
\end{equation}
Here we see clearly the two contributions to the Lyapunov exponent:
although the derivative $f'(x)$ does not change when including noise
in the trajectory, the stationary probability does change (see
Fig.3), thus producing the observed change in the Lyapunov exponents.
Synchronization, then, can be a general feature in maps which have a
large region in which the derivative $|f'(x)|$ is smaller than one. Noise
will be able, then , to explore that region and yield, on the average,
a negative Lyapunov exponent.

The second system we have studied is the well known Lorenz model with
random terms added\cite{lorenz,mar94}:
\begin{eqnarray}
\dot x & = & p(y-x) \nonumber \\
\dot y & = & -x z + r x -y +\epsilon \xi  \label{eq:lor}\\
\dot z & = & x y -b z \nonumber
\end{eqnarray}
$\xi$ now is white noise: a Gaussian random process of mean zero and
delta correlated, $\langle \xi(t) \xi(t') \rangle = \delta (t-t')$.
We have used $p=10$, $b=8/3$ and $r=28$ which, in the deterministic
case, $\epsilon=0$ are known to lead to a chaotic behavior (the largest Lyapunov
exponent is $\lambda \approx  0.9  >0$). We have integrated numerically
the above equations using the Euler method with a time step $\Delta t =
0.001$. For the deterministic case, trajectories starting with different
initial conditions are completely uncorrelated, see Fig. (4a). This
is also the situations for small values of $\epsilon$.
However, when
using a noise intensity $\epsilon=40$ the noise is strong enough
to induce synchronization of the trajectories. Again the presence
of the noise terms makes the largest Lyapunov exponent become
negative (for $\epsilon=40$ it is $\lambda \approx -0.2$).
As in the example
of the map, after some transient time, two different evolutions
which have started in completely different initial conditions synchronize
towards the same value of the three variables (see Fig. (4b) for
the $z$ coordinate). One could argue that the intensity of the noise
is very large. However, the basic structure of the ``butterfly" Lorenz
attractor remains unchanged as shown in Fig. (5).

In conclusion, we have shown that it is possible for noise to
synchronize trajectories of a system which, deterministically, is
chaotic. The novelty of our results is that the noise used in the
two examples, a 1-d map and the Lorenz system, is unbiased, i.e.
has always zero mean.

There still remain many open questions in this
field. They involve the development of a general theory, probably based
in the invariant measure, that could give us a criterion to determine
the range of parameters (including noise levels) for which the Lyapunov
exponent becomes negative, thus allowing synchronization. Another important
question relates the structural stability of this phenomenon. Any
practical realization of this system can not produce two {\sl
identical} samples. If one wants to use sthocastic synchronization
of electronic
emitters and receivers (as a means of encryption) one should
be able to determine which is the allowed discrepancy between
circuits before the lack of synchronization becomes unacceptable.

{\bf Acknowledgements} We thank financial support from DGESIC (Spain)
projects numbers PB94-1167 and PB97-0141-C02-01.

%\newpage

\section*{Figure captions}

%\begin{figure}
%\vspace{0.5 truecm}
%\begin{minipage}{0.5\textwidth}
%\includegraphics[width=8.0cm]{fig1a.ps}
%\end{minipage}\hfill
%\begin{minipage}{0.5\textwidth}
%\includegraphics[width=8.0cm]{fig1b.ps}
%\end{minipage}
%\vspace{0.5 truecm}
%\caption{
%\label{fig1}
FIGURE 1: (a) Bifurcation diagram of the map given by
Eqs.(\ref{eq:2}) and (\ref{eq:3}) in the absence of noise terms.
(b) Lyapunov exponent for the noiseless map ($\epsilon=0$,
continuous line) and the map with a noise intensity $\epsilon=0.1$
(dotted line) and $\epsilon=0.2$ (dot-dashed line).%}
%\end{figure}

%\begin{figure}
%\vspace{0.5 truecm}
%\begin{minipage}{0.5\textwidth}
%\includegraphics[width=8.0cm]{fig2a.ps}
%\end{minipage}\hfill
%\begin{minipage}{0.5\textwidth}
%\includegraphics[width=8.0cm]{fig2b.ps}
%\end{minipage}
%\vspace{0.5 truecm}
%\caption{
%\label{fig2}
FIGURE 2: Plot of two realizations $x^{(1)}$, $x^{(2)}$ of of the
map given by Eqs. (\ref{eq:2}) and (\ref{eq:3}). Each realization
consists of 10,000 points which have been obtained by iteration of
the map starting in each case from a different initial condition
(100,000 initial iterations have been discarded and are not
shown). In figure (a) there is no noise, $\epsilon=0$ and the
trajectories are independent of each other. In figure (b) we have
use a level of noise $\epsilon=0.2$ producing a perfect
synchronization (after discarding some initial iterations). %}
%\end{figure}

%\begin{figure}
%\vspace{0.5 truecm}
%\begin{minipage}{0.5\textwidth}
%\includegraphics[width=8.0cm]{fig3a.ps}
%\end{minipage}\hfill
%\begin{minipage}{0.5\textwidth}
%\includegraphics[width=8.0cm]{fig3b.ps}
%\end{minipage}
%\vspace{0.5 truecm}
%\caption{
%\label{fig3}
FIGURE 3: Plot of the stationary distribution for the map given by
Eqs.(\ref{eq:2}) and (\ref{eq:3}) in the (a) deterministic case
$\epsilon=0$, and (b) the case with noise along the trajectory,
$\epsilon=0.2$.%}
%\end{figure}

%\begin{figure}
%\vspace{0.5 truecm}
%\begin{minipage}{0.5\textwidth}
%\includegraphics[width=8.0cm]{fig4a.ps}
%\end{minipage}\hfill
%\begin{minipage}{0.5\textwidth}
%\includegraphics[width=8.0cm]{fig4b.ps}
%\end{minipage}
%\vspace{0.5 truecm}
%\caption{
%\label{fig4}
FIGURE 4: Same than figure (1) for the $z$ variable of the Lorenz
system, Eqs.(\ref{eq:lor}) in the (a) deterministic case
$\epsilon=0$ and (b) $\epsilon=40$. Notice the perfect
synchronization in case (b).%}
%\end{figure}

%\begin{figure}
%\vspace{0.5 truecm}
%\begin{minipage}{0.5\textwidth}
%\includegraphics[width=8.0cm]{fig5a.ps}
%\psfig{file=fig4a.ps,width=6.0cm}
%\end{minipage}\hfill
%\begin{minipage}{0.5\textwidth}
%\includegraphics[width=8.0cm]{fig5b.ps}
%\psfig{file=fig4b.ps,width=6.0cm}
%\end{minipage}
%\vspace{0.5 truecm}
%\caption{
\label{fig5}
``Butterfly" attractor of the Lorenz system in the cases (a) of no
noise $\epsilon=0$ and (b) $\epsilon=40$. %}
%\end{figure}

\end{document}